\documentclass[twocolumn,amsmath,amssymb,pra]{revtex4-1}

\usepackage{graphicx}
\usepackage{color}
\usepackage{dcolumn}
\usepackage{amsmath}
\usepackage{CJK}
\usepackage{subfigure}
\usepackage{sidecap}
\usepackage{dsfont}

\usepackage{mathrsfs}
\usepackage{amsfonts}
\usepackage{indentfirst}
\usepackage{bm}

\usepackage[colorlinks,citecolor=blue]{hyperref}

\newcommand{\be}{\begin{equation}}
\newcommand{\ee}{\end{equation}}
\newcommand{\bey}{\begin{eqnarray}}
\newcommand{\eey}{\end{eqnarray}}
\newcommand{\bw}{\begin{widetext}}
\newcommand{\ew}{\end{widetext}}

\newcommand{\ba}{\begin{array}}
\newcommand{\ea}{\end{array}}
\newcommand{\bi}{\begin{itemize}}
\newcommand{\ei}{\end{itemize}}
\newcommand{\bem}{\begin{enumerate}}
\newcommand{\eem}{\end{enumerate}}

\begin{document}

\title{Characterizing the Lipkin-Meshkov-Glick model excited state quantum phase transition using
  dynamical and statistical properties of the  diagonal entropy}

\author{Qian Wang\footnote{Electronic address: qwang@zjnu.edu.cn}}

\affiliation{Department of Physics, Zhejiang Normal University, Jinhua 321004, China}
\affiliation{CAMTP-Center for Applied Mathematics and Theoretical Physics, University of Maribor, Mladinska 3, SI-2000
Maribor, Slovenia}

\author{Francisco P\'{e}rez-Bernal\footnote{Electronic address: curropb@uhu.es}}

\affiliation{Departamento de Ciencias Integradas y Centro de Estudios Avanzados en F\'{i}sica,
Matem\'{a}ticas y Computaci\'{o}n, Universidad de Huelva, Huelva 21071, Spain}
\affiliation{Instituto Carlos I de F\'{i}sica Te\'{o}rica y Computacional, Universidad de Granada, Granada 18071, Spain}

\begin{abstract}

Using the diagonal entropy, we analyze the dynamical signatures of the
Lipkin-Meshkov-Glick (LMG) model excited-state quantum phase transition (ESQPT).
We first show that the time evolution of the diagonal entropy behaves as an
efficient indicator of the presence of an ESQPT.  We also compute the
probability distribution of the
diagonal entropy values over a certain time interval and we find
that the resulting distribution provides a clear distinction
between the different phases of ESQPT.  Moreover, we observe that the probability
distribution of the diagonal entropy at the ESQPT critical point has a
universal form, well described by a beta distribution, and that
a reliable detection of the ESQPT can be obtained from the diagonal entropy central moments.

\end{abstract}

\date{\today}

\maketitle

\section{introduction}

The notion of excited-state quantum phase transition (ESQPT)
\cite{Cejnar2006,Caprio2008} was first introduced to describe the nonanalytical
properties in excited states of quantum systems and was soon identified, both
theoretically
\cite{Brandes2013,Magnani2014,Bastidas2014,Puebla2016,Armando2016,Ramos2017,Wang2020,Feldmann2020}
and experimentally
\cite{Larese2011,Larese2013,Dietz2013,Rivera2019,TianT2020,Khaloufrivera2020},
in various many-body systems. For a recently published review on the subject,
see Ref.~\cite{Cejnar2021}. Being a generalization of ground-state quantum phase
transitions (QPTs) \cite{Carr2010,Sachdev2011}, ESQPTs are manifested by the
appearance of a singularity in the density of states -or in one of its
derivatives- at a critical energy value, for fixed Hamiltonian parameters
\cite{Stransky2014, Cejnar2021}.  It has been found that ESQPTs play an
important role in several contexts, including quantum decoherence processes
\cite{Relano2008,Perez2009,Qian2019a}, quantum chaos
\cite{Perez2011a,Hirsch2014,Magnani2015,Lobez2016}, and quantum thermodynamics
\cite{Puebla2015,Qian2017}.  Many efforts have been devoted to understanding the
intriguing static
\cite{YuanZi2012,Puebla2014,Stransky2014,Stransky2015,Cejnar2017,Perez2017a,Milan2017,Macek2019}
and dynamic
\cite{Puebla2013,Engelhardt2015,Qian2019b,Kopylov2017,Perez2011,SantosL2015,SantosL2016,
  Hummel2019,Perez2017b,mzaouali2021} properties of this new type of phase
transition.

Motivated by the recent advances on experimental techniques, the study of
nonequilibrium dynamics of isolated quantum systems has received much attention
in the past few years
\cite{Sengupta2011,Langen2015,Bhattacharyya2015,Haldar2020}.  Along this
direction, it is natural and important to explore how ESQPTs influence
nonequilibrium dynamics of isolated systems.  To date, several remarkable
dynamical effects of ESQPTs have been revealed: an enhanced survival probability
decay \cite{Perez2011,SantosL2015,SantosL2016,Perez2017b,Kloc2018}, an
exponential growth of out-of-time-order correlators \cite{Cameo2020}, and
singularities in the time evolution of observables \cite{Engelhardt2015}.
Moreover, the investigation of how to dynamically probe ESQPTs is also under an
active development
\cite{Puebla2013,Perez2017b,Qian2019a,Qian2019b,Feldmann2020,mzaouali2021} and
implies possible ways of experimental exploration of ESQPTs through their
evidences in the dynamics of isolated quantum many-body systems.  In spite of
these many works, quite a few aspects of the dynamical signatures of ESQPTs are
still under discussion and more works are required in order to get a deeper
understanding of the properties of ESQPTs.

In this work, we consider the Lipkin-Meshkov-Glick (LMG) model \cite{Lipkin1965,
  Lipkin1965b, Lipkin1965c} and study the dynamical features of its ESQPT by
means of the diagonal entropy. The diagonal entropy, for a given set of energy
eigenstates, is defined as $\mathcal{S}_d=-\sum_n\rho_{nn}\ln{\rho_{nn}}$, where
$\rho_{nn}$ are the diagonal elements of the density matrix $\rho$ in the basis
of energy eigenstates \cite{Polkovnikov2011}. This definition connects this
quantity with the Shannon information entropy of the probability
distribution corresponding to the energy eigenbasis \cite{mzaouali2021}.  The diagonal entropy
exhibits most of the properties of a thermodynamic entropy, including
additivity. Hence, it remains constant in adiabatic processes and it increases
when systems are taken out of equilibrium.  That makes the diagonal entropy a
fine option for the study of nonequilibrium dynamics in isolated quantum
\cite{Santos2011,Mata2015,Giraud2016,TorresH2017,SunZ2020, ZWang2020}.
Moreover, $\mathcal{S}_d$ is consistent with the well-known von Neumann's
entropy for systems in equilibrium.  It is also worth mentioning that, since the
diagonal entropy only involves the diagonal part of the density matrix, in
principle it can be experimentally accessed \cite{SunZ2020}.

In the present work, we first focus on the time evolution of the diagonal
entropy in a cyclic quench. We show that the time evolution of the diagonal
entropy reveals the ESQPT existence displaying qualitatively distinct dynamics
in the different ESQPT phases.  Then, we investigate the probability
distribution of the diagonal entropy values over a certain time interval. We
show how the underlying ESQPT determines the distribution statistical
properties.  In particular, at the ESQPT critical energy, the diagonal entropy
probability distribution has a universal form, independent of the system size
and the Hamiltonian parameter values, that is in good agreement with the beta
distribution. We also show that it is possible to detect the ESQPT from the
values of the central moments of the diagonal entropy probability distribution.

The article is structured as follows.  In Sec.~\ref{Sec2}, we describe the
protocol used in this work and introduce the LMG model, briefly reviewing its
main properties.  In Sec.~\ref{Sec3}, we present our main results and discuss
how the signatures of ESQPT can be identified in the dynamics of the diagonal
entropy as well as its statistical properties.  Finally, we summarize the main
conclusions of this work in Sec.~\ref{Sec4}.

\section{Protocol and model} \label{Sec2}

\subsection{Protocol and diagonal entropy}

\begin{figure}
   \begin{minipage}{1\linewidth}
    \includegraphics[width=\columnwidth]{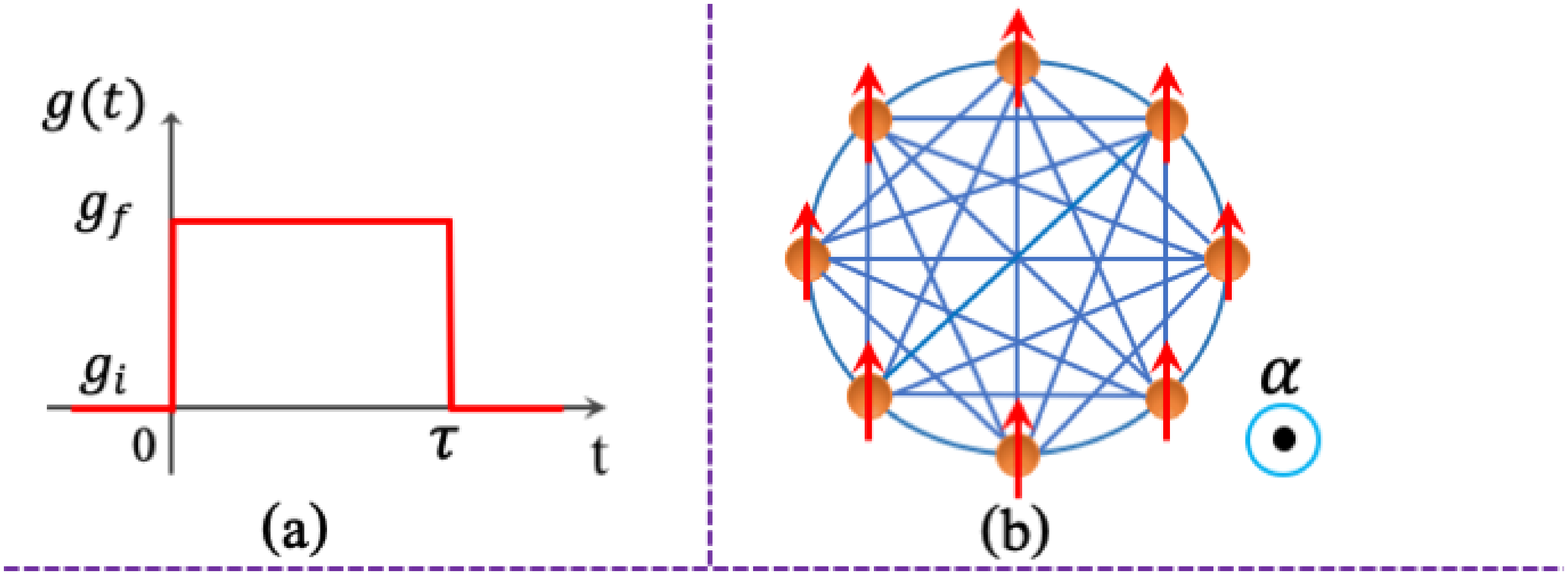}
   \end{minipage}
   \begin{minipage}{1\linewidth}
    \includegraphics[width=\columnwidth]{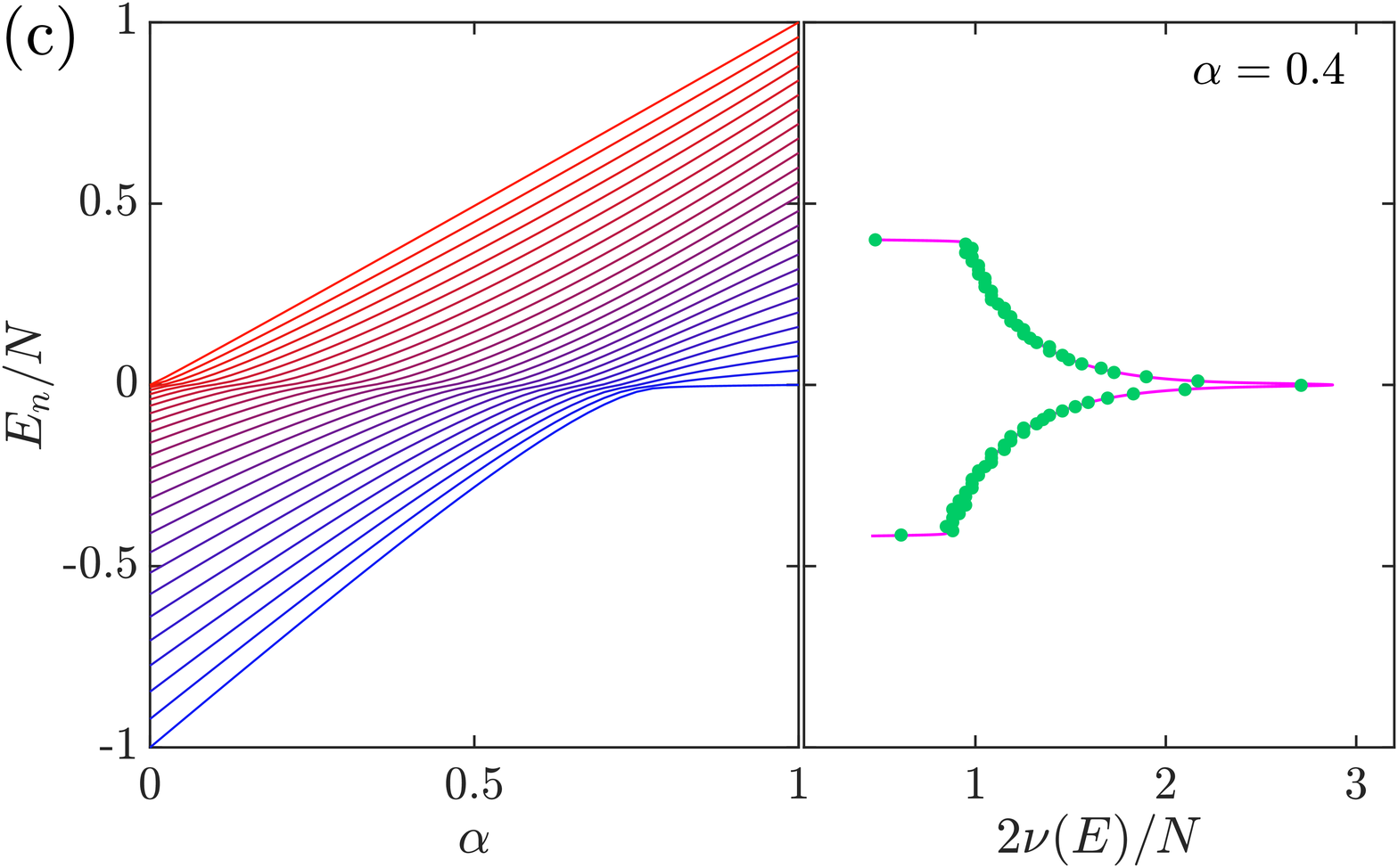}
    \end{minipage}
    \caption{(a) The quench protocol studied in this work, as described in the text. 
    (b) Schematic representation of the LMG model. Spins are fully connected through an infinite range coupling
    and in an external magnetic field with strength $\alpha$ along the $z$ direction.
    (c) Rescaled even-parity energy spectrum of the Lipkin model as a function of $\alpha$ with $N=50$ (left panel) 
    and the rescaled density of states of the LMG model for $\alpha=0.4$ with $N=5000$ (right panel). 
    The green dots denote the numerical results, while the solid line is obtained via Eq.~(\ref{SCDoS}).
    All quantities are dimensionless.} 
    \label{SchmaticF}  
  \end{figure}

Assuming the system under study is described by a Hamiltonian $H(g)$, with $g$
being a control parameter, we consider a cycle protocol with sudden changes of
the control parameter at two different times.  As depicted in
Fig.~\ref{SchmaticF}(a), the protocol consists of the following processes.  (i)
Initially, the control parameter value is $g_i$, the Hamilitonian is $H_i$, and
the system is in the state
$\rho^{(i)}_n=|\psi^{(i)}_n\rangle\langle\psi^{(i)}_n|$, where
$|\psi^{(i)}_n\rangle$ is the $n$-th $H_i$ eigenstate, with eigenvalue
$E_n^{(i)}$.  (ii) At time $t=0$, the control parameter is suddenly changed
(quenched) from the initial value $g_i$ to a final value $g_f$ and the
Hamiltonian of the system is a new one, $H_f$, with eigenstates
$|\psi_n^{(f)}\rangle$ and eigenvalues $E_n^{(f)}$.  From $t=0$ on, the dynamics
of the system is governed by the Hamiltonian $H_f$.  (iii) At time $t=\tau$, the
system undergoes a second quench, which changes the control parameter from $g_f$
back to its initial value $g_i$, completing the cycle protocol.  From now on, the
system evolves under $H_i$ for $t\geq\tau$.

The state of the system at $t=\tau$ is given by
$\rho_\tau=e^{-iH_f\tau}\rho_n^{(i)}e^{iH_f\tau}$ and, therefore, the diagonal
entropy at \(t = \tau\) in the basis of eigenstates of the $H_i$ Hamiltonian can
be written as \be \label{SdtE} \mathcal{S}_d(\tau) =
-\sum_k\mathcal{C}_k(\tau)\ln\mathcal{C}_k(\tau), \ee where
$\mathcal{C}_k(\tau)=|\langle \psi_k^{(i)}|e^{-iH_f\tau}|\psi_n^{(i)}\rangle|^2$
and $|\psi_k^{(i)}\rangle$ is the $k$-th eigenstates of Hamiltonian $H_i$
\cite{Mata2015}.  As already mentioned, it has been argued that the diagonal
entropy fulfills the second law of thermodynamics, namely, it grows when a
system is taken out of equilibrium, it saturates at the equilibration time
scale, it is an additive quantity, and it is conserved for adiabatic processes
\cite{Polkovnikov2011, Santos2011}.  Note that $\mathcal{C}_k(\tau)$ is equal to
the well-known survival probability when we take $k=n$ and that, independently
of $\tau$ and $H_f$ values, $\sum_k\mathcal{C}_k(\tau)=1$.

The diagonal entropy is a  non-linear function of the density
matrix and, therefore, the long-time averaged diagonal entropy, denoted as $\overline{\mathcal{S}_d(\tau)}$, 
is not equal to  the diagonal entropy for the long-time averaged state, 
$\langle \mathcal{S}_d\rangle$ \cite{Ikeda2015,Mata2015}. 
It has been conjectured that for a pure initial state, the deviation between 
these two quantities, \(\Delta\), satisfies the inequality
$\Delta=\langle \mathcal{S}_d\rangle-\overline{\mathcal{S}_d(\tau)}\leq 1-\gamma$,
where $\gamma=0.5772\dots$ is the Euler's constant \cite{Ikeda2015}.
As $\Delta$ fluctuations are  minimal  once the system is in equilibrium, it has been employed to 
explore the connection between relaxation and transitions between integrability and chaos in 
various quantum systems \cite{Mata2015,Giraud2016}. 
In the present work, we pay heed to the ESQPT signatures  in the nonequilibrium dynamics of a quantum isolated system, 
investigating the dynamical and statistical properties of the diagonal entropy of 
the LMG model, in which the above mentioned cycle protocol is implemented.

  \begin{figure}
    \includegraphics[width=\columnwidth]{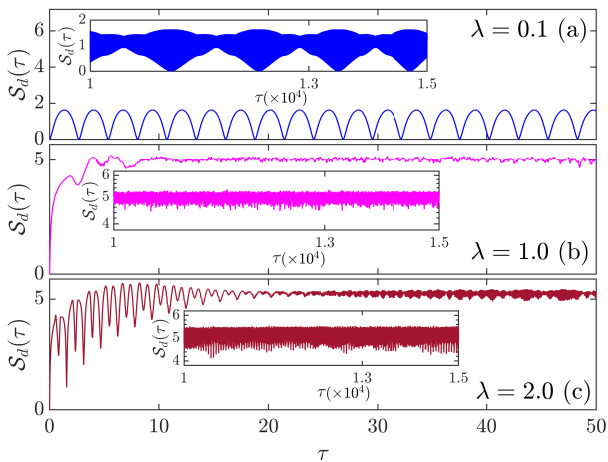}
    \caption{Time evolution of $\mathcal{S}_d(\tau)$ for three different values of 
    $\lambda$, with control parameter $\alpha=0.4$ and system size $N=1000$.
    The inset in each panel shows the long time behavior of $\mathcal{S}_d(\tau)$ with 
    $\tau$ in a range from $10^4$ to $1.5\times10^4$ for the corresponding
    $\lambda$ values.
    The axes in all figures are dimensionless.} 
    \label{TSd}  
  \end{figure}

\subsection{Lipkin-Meshkov-Glick (LMG) model}  

The LMG model, originally introduced as a toy model in nuclear physics
\cite{Lipkin1965, Lipkin1965b, Lipkin1965c}, was later found to be useful in
many areas of physics
\cite{Ribeiro2007,Puebla2015,Santos2015,Campbell2015,Russomanno2017,Cameo2020,Tianrui2020}  
and has been realized with high precision in different experimental platforms \cite{Albiez2005,Zibold2010,Leroux2010,Makhalov2019}.
In particular, it has been used as a  paradigmatic model in the study of ESQPTs
\cite{Caprio2008,Relano2008,Perez2009,YuanZi2012,SantosL2016,Qian2017,Qian2019a,Qian2019b}. 
This model can be mapped to the transverse Ising model with infinite-range
interactions. Hence, the LMG model describes $N$ fully connected $1/2$-spin particles coupled to an external transverse field with strength $\alpha$; 
see Fig.~\ref{SchmaticF}(b) for a schematic representation of the LMG model.

Employing the collective spin operators $J_\beta=\sum_l\sigma_\beta^l/2$, where $\beta=\{x,y,z\}$ 
and $\sigma^l_\beta$ are Pauli spin matrices for the \(l\)-th spin, the Hamiltonian of the LMG model can be written as 
\be \label{LMGH}
  H=-\frac{4(1-\alpha)}{N}J_x^2+\alpha\left(J_z+\frac{N}{2}\right),
\ee
where $N$ is the total number of spins and the control parameter  $\alpha\in[0,1]$ is the strength of the magnetic field along 
the $z$ direction. For simplicity's sake, we consider $\hbar=1$ throughout this work and set the quantities 
studied in this article as dimensionless. 
 
The Hamiltonian in Eq.~(\ref{LMGH}) conserves the total spin $\mathbf{J}^2=J_x^2+J_y^2+J_z^2$, 
whose eigenvalues are $j(j+1)$ with $0\leq j\leq N/2$.
We perform our calculations in the sector of maximum angular momentum, $j=N/2$, with dimension $N+1$. 
Moreover, as the parity operator $\Pi=e^{i\pi(J_z+j)}$ also commutes with $H$, 
the Hamiltonian matrix in $j=N/2$ sector can be further split into two blocks, an even parity block, with dimension $N/2+1$,
and an odd parity block, with dimension $N/2$.
We further restrict our calculations to the even parity block, which includes
the system ground state. 

The elements of the Hamiltonian matrix in the basis of eigenstates of $J_z$,  $|j,m_z\rangle$, 
with $-N/2\leq m_z\leq N/2$, are given by
\begin{widetext}
\begin{align}
  &\langle j,m_z|H| j,m_z\rangle=q_\alpha\left(\frac{N}{2}+m_z\right)+\alpha-1, \notag \\
  &\langle j,m_z+2|H| j,m_z\rangle=-\frac{1-\alpha}{N}\sqrt{\left(\frac{N}{2}-m_z-1\right)}
         \sqrt{\left(\frac{N}{2}-m_z\right)\left(\frac{N}{2}+m_z+1\right)\left(\frac{N}{2}+m_z+2\right)}, \notag
\end{align}
\end{widetext}
where $q_\alpha=[2(1-\alpha)m_z/N]+2\alpha-1$.

  \begin{figure*}
    \includegraphics[width=\textwidth]{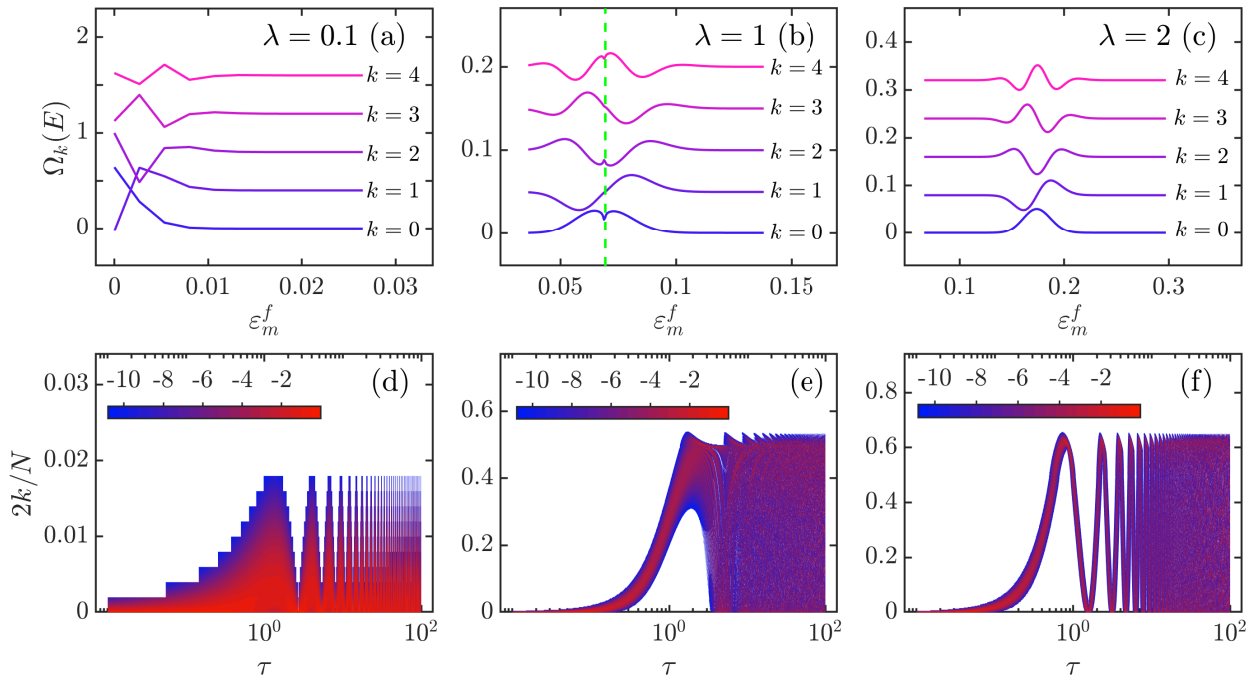}
    \caption{Panels (a)-(c): $\Omega_k(E)$ as a function of the rescaled energy
      for the first $5$ eigenstates ($k=0,1,\ldots,4$) with $\lambda=0.1,1,2$.  The rescaled energy is defined as
      $\varepsilon_m^f=[E_m^{(f)}-E_0^{(f)}]/[E_{max}^{(f)}-E_0^{(f)}]$, where $E_0^{(f)}$ is the
      ground state energy of $H_f$, while $E_{max}^{(f)}$ denotes the maximum energy
      of $H_f$. To offer a three-dimensional-like visualization, the different $\Omega_k(E)$
      curves are shifted in the $y$-direction
      by $0.4k$ (a), $0.05k$ (b), and $0.08k$ (c). 
      The green dashed line in panel (b) indicates the rescaled ESQPT  critical energy.
      Panels (d)-(f): Heat map plot depicting $\ln[\mathcal{C}_k(\tau)]$ as a function of $k$ and $\tau$ for the same values
      of $\lambda$ as in panels (a)-(c). 
      White color indicates $\mathcal{C}_k(\tau)=0$.
      In all cases the control parameter $\alpha=0.4$ and the system size is $N=1000$.
      All quantities are dimensionless.} 
    \label{Oct}  
  \end{figure*}

The LMG Hamiltonian in Eq.~(\ref{LMGH}) undergoes a second-order ground state quantum phase transition 
at the critical point $\alpha_c=0.8$ \cite{Romera2014,Castanos2018}. The system
is in the broken-symmetry phase when $\alpha<\alpha_c$ and in the symmetric phase
for $\alpha\ge\alpha_c$.
Another remarkable feature of the LMG model is the occurrence of an 
ESQPT for $\alpha<\alpha_c$ \cite{Caprio2008,Relano2008,Perez2009,Hummel2019}.
ESQPTs in systems with a single effective degree of freedom, like the LMG model,
are characterized by a high density of excited levels at a critical energy
value, $E_c$. The level density is nonanalytical in the mean field limit (large \(N\)
limit) of the system \cite{Cejnar2021}. This is illustrated for the LMG model in the left panel of
Fig.~\ref{SchmaticF}(c), where it is clear how energy levels are
piling up the neighborhood of the critical energy $E_c=0$.
 
The eigenvalues clustering at $E_c=0$ leads to a cusp singularity in 
the density of states, $\nu(E)$, defined as $\nu(E)=\sum_n\delta(E-E_n)$.
In the semiclassical limit $N\to\infty$, $\nu(E)$ can be analytically calculated as \cite{Perez2009,Qian2019a}
\be \label{SCDoS}
  \nu(E)=\frac{N}{2\pi}\int\delta[E-\mathcal{H}_{cl}(x,p)] dxdp,
\ee
where $\mathcal{H}_{cl}$ is the classical counterpart of $H$ in Eq.~(\ref{LMGH}).
The right panel of Fig.~\ref{SchmaticF}(c) plots the density of states for the case of  $\alpha=0.4$ with $N=5000$.
We observe that $\nu(E)$ obtained by means of Eq.~(\ref{SCDoS}) has an excellent
agreement with the numerical data and it is evident the expected cusp divergence at $E_c=0$.
In the following, we focus on the identification of the signatures of this ESQPT in the 
dynamical and statistical properties of the diagonal entropy. 

  \begin{figure*}
    \includegraphics[width=\textwidth]{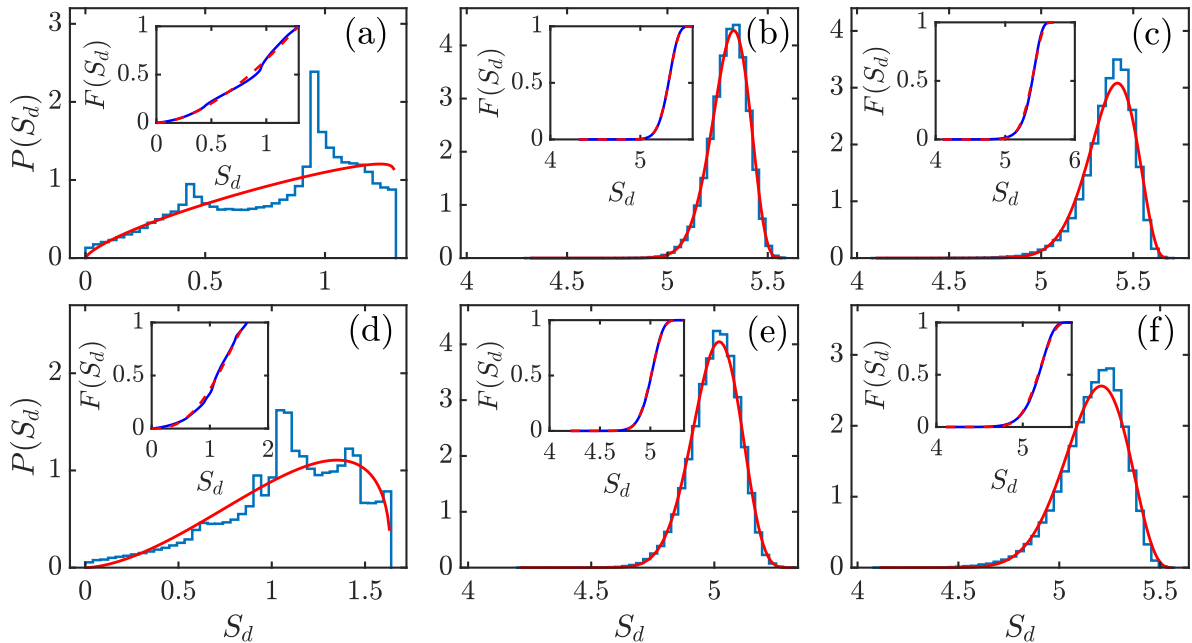}
    \caption{Panels (a)-(c): Probability distribution of the diagonal entropy
      $P(S_d)$ for
    $\lambda=0.1$ (a), $\lambda=\lambda_c^\alpha=1.5$ (b), and $\lambda=2$
      (c). The control parameter is fixed to $\alpha=0.2$.
    Panels (d)-(f): Probability distribution of the diagonal entropy $P(S_d)$ for $\lambda=0.1$ (a),
    $\lambda=\lambda_c^\alpha=1$ (b), and $\lambda=2$ with $\alpha=0.4$. 
    In all cases the system size is $N=1000$.
    The red solid line in each panel denotes the corresponding optimized beta distribution from Eq.~(\ref{BtPdf}).
    The  parameter values ($a,b,S_0,S_m$) are 
    (a) $(1.722,1.038,0,1.295)$, (b) $(23.705,6.637,4.316,5.582)$, (c) $(18.172,4.638,4.101,5.692)$,
    (d) $(2.697,1.362,0,1.64)$, (e) $(18.972,7.964,4.208,5.334)$, (f) $(12.292,4.676,4.109,5.567)$.
    The inset in each panel shows with a blue solid line the cumulative distribution function of the diagonal entropy \(F(S_d)\)
    and the corresponding result for the beta distribution (red dashed line), respectively. 
    All quantities are dimensionless.} 
    \label{Prb}  
  \end{figure*}

\section{The LMG model diagonal entropy} \label{Sec3}

In the first hand, we focus on the dynamics of the diagonal entropy $\mathcal{S}_d(\tau)$, 
and in the second hand, we consider the distribution of values of $\mathcal{S}_d(\tau)$ with $\tau\geq0$.
We are mainly interested in how the ESQPT affects
the time evolution of $\mathcal{S}_d(\tau)$ and 
the $\mathcal{S}_d(\tau)$ probability distribution, as well as the moments of this distribution.

In our study, the above described cycle protocol is achieved as follows.
Initially, the system is at the ground state, $|\psi_0^{(i)}\rangle$, of
Hamiltonian (\ref{LMGH}) with $H_i=H$, $g_i=0$, and
$\rho_0^{(i)}=|\psi_0^{(i)}\rangle\langle\psi_0^{(i)}|$.  At time $t=0$, we turn
on an external magnetic field along the $z$ direction with strength $\lambda$.
We thus have $g_f=\lambda$ and $H_f=H+\lambda(J_z+N/2)$.  The external magnetic
field is then switched off at time $t=\tau$ to back to the starting point,
completing the closed cycle.  The diagonal entropy at time $t=\tau$,
$\mathcal{S}_d(\tau)$, is given by Eq.~(\ref{SdtE}) with
\be \label{Cfk}
\mathcal{C}_k(\tau)=|\langle\psi_k^{(i)}|e^{-iH_f\tau}|\psi_0^{(i)}\rangle|^2=\left|\int
dE\Omega_k(E)e^{-iE\tau}\right|^2.  \ee
\noindent Here, $|\psi_k^{(i)}\rangle$ is the $k$th eigenstate of $H$ in
Eq.~(\ref{LMGH}) and
\be \label{OmgD}
\Omega_k(E)=\sum_m\langle\psi_k^{(i)}|\psi_m^{(f)}\rangle\langle
\psi_m^{(f)}|\psi_0^{(i)}\rangle\delta[E-E_m^{(f)}],
\ee
\noindent with $|\psi_m^{(f)}\rangle$ denotes the $m$th eigenstate of $H_f$
corresponding to the eigenvalue $E_m^{(f)}$. We point out that results
qualitatively similar to the reported ones are obtained for different choices of
the initial state.

The system can be driven through the critical energy of ESQPT by varying the
strength of the external magnetic field,  $\lambda$. We define the critical strength, denoted as $\lambda_c^\alpha$, 
as the magnetic field intensity that brings the system, initially in the ground state, 
to the critical energy, $E_c=0$. In the LMG model case this critical strength can be obtained 
using the semiclassical approach \cite{Relano2008,Perez2009}
\be \label{CriticalV}
   \lambda_c^\alpha=\frac{1}{2}\left(4-5\alpha\right),
\ee
where $\alpha\in(0,4/5)$.
We would like to point out that the ESQPT critical strength, $\lambda_c^\alpha$,
differs from the critical strength for the ground state quantum phase
transition, $\lambda_{c0}^\alpha$\cite{Perez2009}.

  \begin{figure}
    \includegraphics[width=\columnwidth]{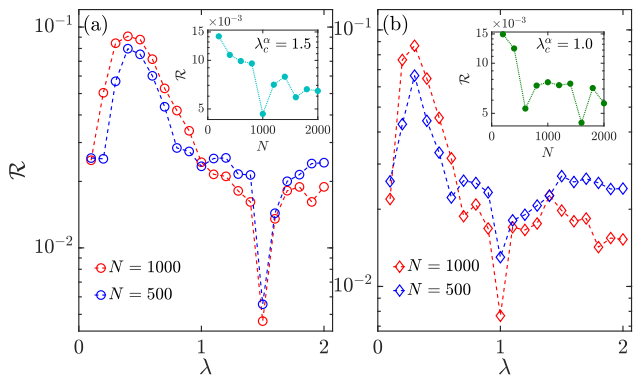}
    \caption{$\mathcal{R}$ in Eq.~(\ref{Drmse}) as a function of $\lambda$ for different system size
      $N=500$ and $1000$,  with control parameter values $\alpha=0.2$ (a) and $\alpha=0.4$ (b).
    The inset in (a) shows $\mathcal{R}$ as a function of system size $N$ for $\alpha=0.2$ and $\lambda_c^\alpha=1.5$,
    while the inset in (b) plots $\mathcal{R}$ as a function of $N$ for $\alpha=0.4$ and $\lambda_c^\alpha=1.0$. 
    The axes in all figures are dimensionless.} 
    \label{Rmse}  
  \end{figure}

\subsection{Dynamical behavior of $\mathcal{S}_d(\tau)$}

As a starting point, we investigate the signatures of the ESQPT in the dynamics
of the LMG diagonal entropy.  In Fig.~\ref{TSd}, the diagonal entropy is
depicted as a function of \(\tau\), $\mathcal{S}_d(\tau)$, for three different
values of $\lambda$. In all cases the control parameter $\alpha=0.4$ and the
system size $N=1000$.  In this case, according to Eq.~(\ref{CriticalV}), we have
$\lambda_c^\alpha=1$.  From Fig.~\ref{TSd}, it is clear that the behavior of
$\mathcal{S}_d(\tau)$ as a function of $\tau$ strongly depends on the $\lambda$
value.  Specifically, for $\lambda<\lambda_c^\alpha$, $\mathcal{S}_d(\tau)$
periodically oscillates around a small value, as shown in Fig.~\ref{TSd}(a).
Increasing $\lambda$ leads to an increase in the $\mathcal{S}_d(\tau)$ value
while the initially regular oscillations gradually change towards an irregular
pattern.  As can be seen from Fig.~\ref{TSd}(b), once
$\lambda=\lambda_c^\alpha=1$, $\mathcal{S}_d(\tau)$ displays a fast growth which
rapidly saturates at a maximum value with tiny fluctuations. Notice that the
suppression of the oscillating behavior has also been found in the survival
probability dynamics \cite{Kloc2018}.  As it is shown in
Refs.~\cite{Kloc2018,Hernndez2018}, this feature stems from the fact that
eigenstates having different structure are dynamically entangled at the ESQPT
critical energy.  Above the critical point, e.g.~ the $\lambda=2$ case depicted
in Fig.~\ref{TSd}(c), we observe that $\mathcal{S}_d(\tau)$ increases with time,
with larger oscillations, until it irregularly oscillates around the same
saturation value as in the previous case.

The observed features in the dynamics of the diagonal entropy indicate that the
underlying system ESQPT has a strong impact on the equilibration process of the
quenched system.  Obviously, these features can be used to detect the existence
of an ESQPT, through the singular behavior of $\mathcal{S}_d(\tau)$ at
$\lambda=\lambda_c^\alpha$.  Moreover, different phases of an ESQPT can also be
identified by the distinct behaviors of the diagonal entropy for
$\lambda<\lambda_c^\alpha$ and $\lambda>\lambda_c^\alpha$, respectively.

  \begin{figure}
    \includegraphics[width=\columnwidth]{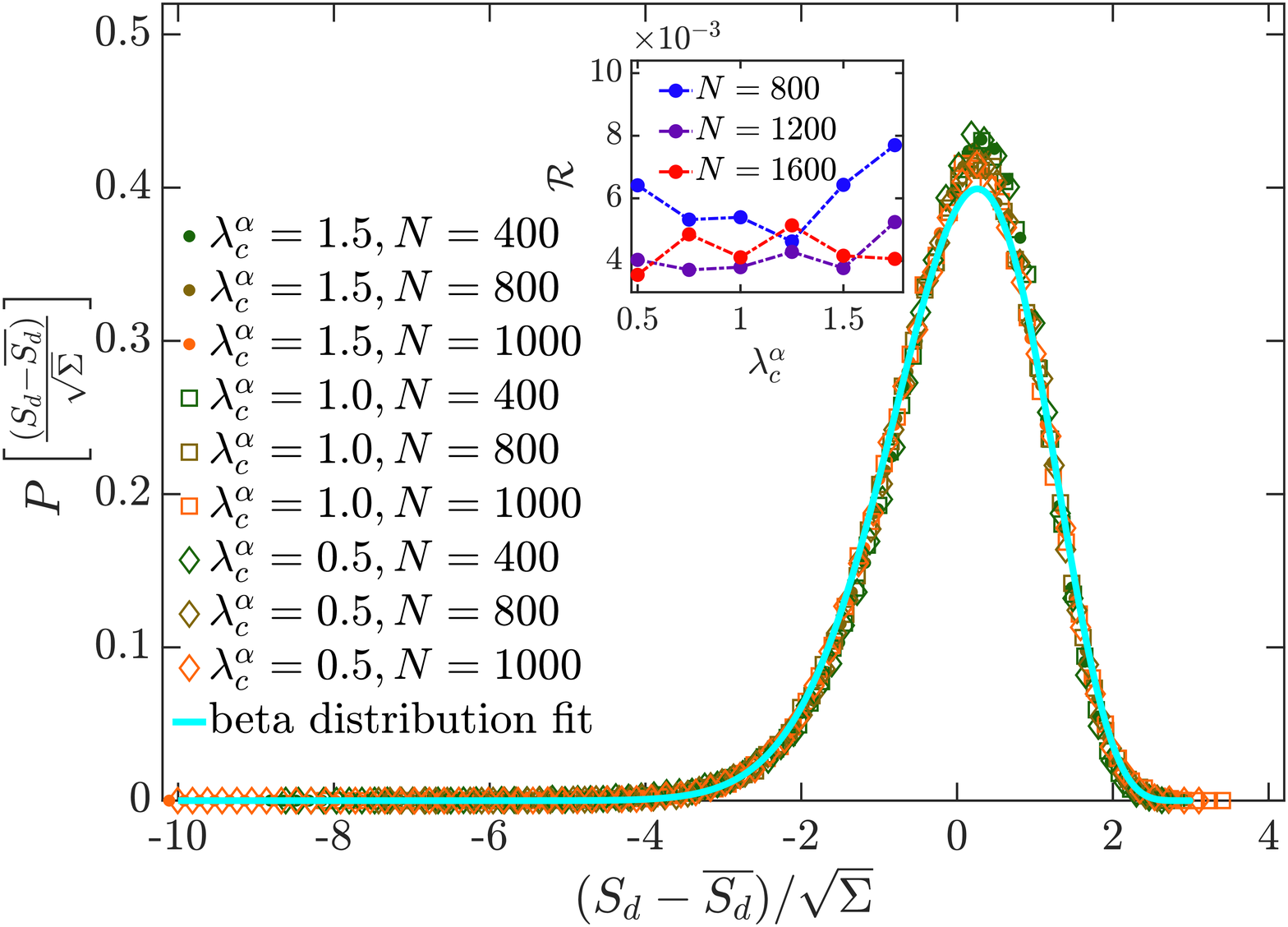}
    \caption{Probability distribution of the shifted and rescaled diagonal entropy, i.~e., 
    $(S_d-\overline{S_d})/\sqrt{\Sigma}$, at different $\lambda_c^\alpha$ and for several system sizes, $N$.
    Here, $\overline{S_d}$ denotes the averaged $S_d$, and $\Sigma$ is the variance of $S_d$.
    The cyan solid line denotes the fitted beta distribution [cf.~Eq.~(\ref{BtPdf})] 
    with fitting parameters $(a,b,S_0,S_m)$ given by $(22.308,6.692,-10,3)$.
    Inset: The RMSE, $\mathcal{R}$, between $P[(S_d-\overline{S_d})/\sqrt{\Sigma}]$ and the fitted beta distribution 
    as a function of $\lambda_c^\alpha$ for different system sizes, $N$. 
    The axes in all figures are dimensionless.} 
    \label{Updf}  
  \end{figure}

To understand the features exhibited by $\mathcal{S}_d(\tau)$, 
we note that, as indicated in Eq.~(\ref{Cfk}), $\mathcal{C}_k(\tau)$ is
the square modulus of the Fourier transform of $\Omega_k(E)$, defined in Eq.~(\ref{OmgD}). 
Therefore, the remarkably different dependence of \(\tau\) for the 
$\mathcal{S}_d(\tau)$ depicted in the different panels of Fig.~\ref{TSd} stems from
the change of  the  $\Omega_k(E)$ properties  as the system 
straddles through the ESQPT.
The behavior of
$\mathcal{S}_d(\tau)$ at the critical energy of the ESQPT can be explained from
the $\Omega_k(E)$ singular structure.
To cast light upon this particular point,
we plot $\Omega_k(E)$ and the corresponding $\mathcal{C}_k(\tau)$ in
Fig.~\ref{Oct}. For the sake of comparison use the same
$\lambda$ values than in  Fig.~\ref{TSd} and, again, 
a control parameter value $\alpha=0.4$ and a system size $N=1000$. 

For the case of $\lambda=0.1<\lambda_c^\alpha$, as shown in Fig.~\ref{Oct}(a),
nonzero $\Omega_k(E)$ values are rather localized at low $H_f$ eigenenergies and
the main contribution is due to states with $k\leq3$.  The simple structure of
$\Omega_k(E)$ in this case explains the oscillations in \(\tau\) of
$\mathcal{C}_k(\tau)$, that occur for small $k$ values, with
$\mathcal{C}_k(\tau)=0$ for other values of $k$, as it is illustrated in
Fig.~\ref{Oct}(d).  This implies that $\mathcal{S}_d(\tau)$ is a periodic
function of  \(\tau\) as shown in Fig.~\ref{TSd}(a).  As $\lambda$
increases, the number of states contributing to $\Omega_k(E)$ increases,
involving states with larger \(k\) values. This, in turn, involves an increase in
the $\mathcal{S}_d(\tau)$ value. Once the critical point
$\lambda=\lambda_c^\alpha=1$ is explored, we obtain the results plotted in
Fig.~\ref{Oct}(b). As the involved values of $k$ are larger, the
complexity of $\Omega_k(E)$ increases.  However, in this particular case, a most
remarkable feature of
$\Omega_k(E)$ is the cusp-like shape near the ESQPT critical energy (marked in
Fig.~\ref{Oct}(b) by a
light green dashed line), occurring at all $k$ values.  As shown in
Ref.~\cite{Perez2011}, the same cusp-like structure in $\Omega_0(E)$ leads to a
fast decay of the survival probability,
$\mathcal{C}_0(\tau)$, followed by random
oscillations with tiny amplitude.  In the present work, we find that cusps
in $\Omega_k(E)$ for nonzero $k$ have the same effect on the time evolution of the
correspondent $\mathcal{C}_k(\tau)$, as illustrated in Fig.~\ref{Oct}(e).
Therefore, the behavior of $\mathcal{S}_d(\tau)$ at $\lambda=\lambda_c^\alpha$
can be traced back to the cusp-like structures in $\Omega_k(E)$ at the critical
energy of ESQPT.  When $\lambda=2>\lambda_c^\alpha$, the structure of
$\Omega_k(E)$ at small values of $k$ are regular, whereas as $k$ increases
the structures of $\Omega_k(E)$ become more and more complex, in a similar way
to the case in Fig.~\ref{Oct}(d).  As a consequence, the  behavior
of $\mathcal{C}_k(\tau)$ is initially regular, followed by small irregular oscillations with
$\mathcal{C}_k(\tau)\approx0$ [see Fig.~\ref{Oct}(f)].  This explains the slow
growth of $\mathcal{S}_d(t)$ at times close to zero for the $\lambda=2$ case
[see Fig.~\ref{TSd}(c)].

These results strongly indicate that the LMG model ESQPT 
has a very significant impact on the equilibration processes that follow a
quench. Therefore, the time 
dependent behavior of the diagonal entropy can be used to reliably distinguish
among the different phases of the ESQPT.   
In addition to this, at the critical point, the particular dynamical behavior of the diagonal entropy acts as a good indicator of 
the presence of ESQPT.

  \begin{figure}
    \includegraphics[width=\columnwidth]{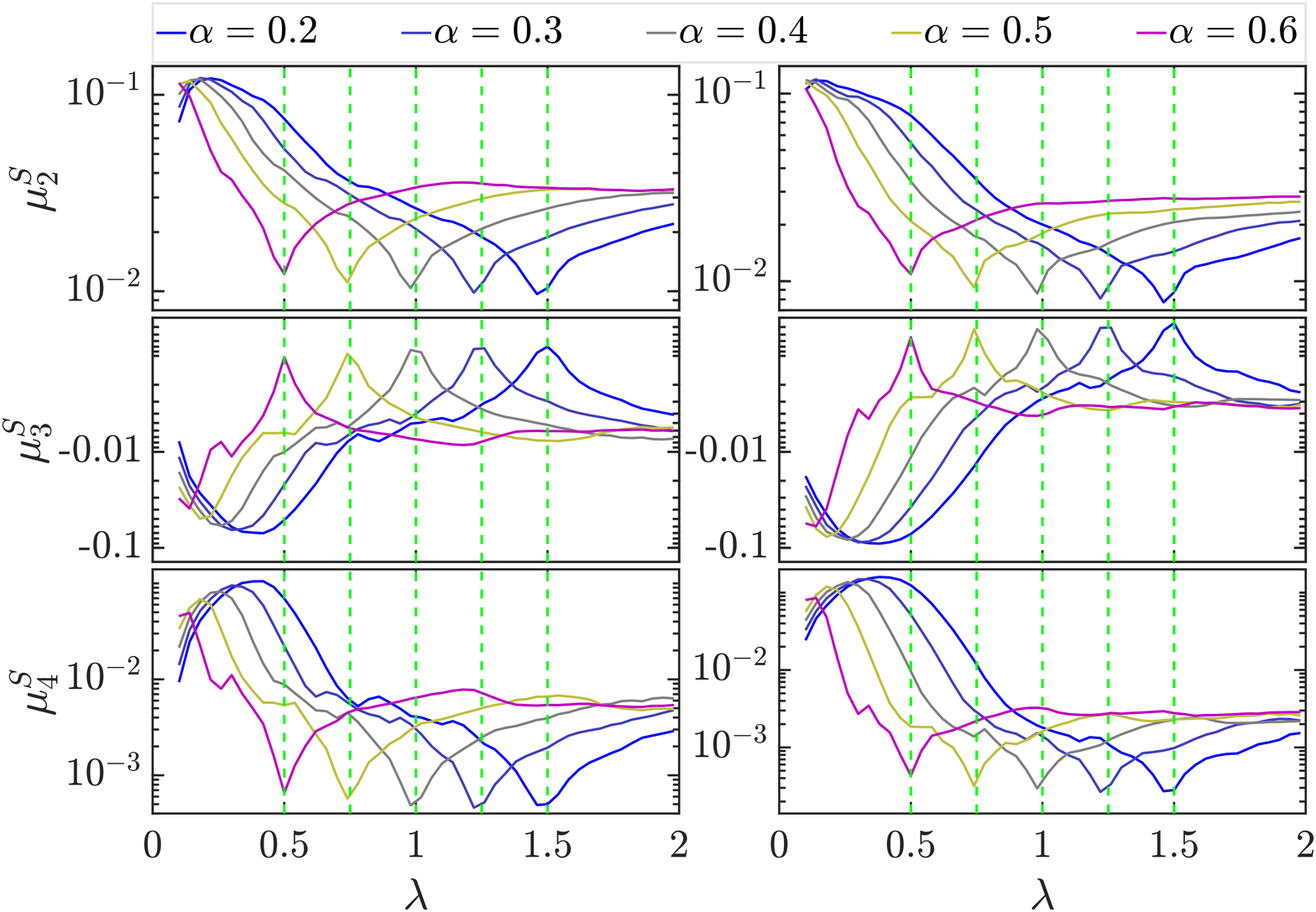}
    \caption{Second, third, and fourth central moments (as labelled) of $P(S_d)$ 
    as a function of $\lambda$
    for $\alpha = 0.2, 0.3,\ldots, 0.6$ and a system size value $N=500$ (left column) and $N=1000$ (right column). 
    The vertical dashed lines in each panel mark the critical values $\lambda_c^\alpha$ for each corresponding $\alpha$.
    The axes in all figures are dimensionless.} 
    \label{CentralM}  
  \end{figure}

\subsection{Statistical properties of $\mathcal{S}_d(\tau)$}

In this subsection we explore the statistical properties of the diagonal
 entropy to gain further insight on how the ESQPT influences the nonequilibrium dynamics
of the LMG model after the quantum quench.
To this end, we investigate the distribution of values of the diagonal entropy
in a long-time interval, considering the probability distribution of
$\mathcal{S}_d(\tau)$ values in a time window $[\tau_0,\tau_0+\Delta\tau]$
\be \label{PdfSd}
   P(S_d)=\lim_{\Delta\tau\to\infty}\frac{1}{\Delta\tau}\int_{\tau_0}^{\tau_0+\Delta\tau}
            \delta[\mathcal{S}_d(\tau)-S_d]d\tau,
            \ee
\noindent where the value of $\tau_0$ is much larger than the initial time scale. 
The correct calculation of  this distribution function implies the consideration
of all the $\mathcal{S}_d(\tau)$  intricacies 
(see, e.~g.~the insets in Fig.~\ref{TSd}); 
which means that we need to evolve the system for a long period of time. 
In our simulation, we take $\tau_0=\Delta\tau=10^4$.
We have carefully checked that the results obtained for larger $\tau_0$ and $\Delta\tau$ values
 do not modify the present conclusions.
The cumulative distribution function of $S_d$ is given by
\be \label{SdCdf}
    F(S_d)=\int_{S_0}^{S_d}P(x)dx,
\ee
where $S_0$ is the minimal value of the distribution range and $P(x)$ is the 
probability distribution function in Eq.~(\ref{PdfSd}).

In Fig.~\ref{Prb}, we plot the computed probability distribution of the diagonal entropy
and the corresponding cumulative distribution for two values of the control
parameter, $\alpha=0.2$ [panels (a)-(c)] and $\alpha=0.4$ [panels (d)-(f)] with
a system size, $N=1000$. In both cases, we include values of $\lambda$ below, at,
and above the critical value $\lambda_c^\alpha$.  From this figure it can be
observed that $P(S_d)$ is a doubly peaked distribution at low $\lambda$ values,
due to the periodic oscillations in $\mathcal{S}_d(\tau)$.  Meanwhile, the small
amplitude of the $\mathcal{S}_d(\tau)$ oscillations is translated to nonzero
values of $P(S_d)$ at low values of $S_d$.  As $\lambda$ value increases, the
growing in $\mathcal{S}_d(\tau)$ shifts $P(S_d)$ towards higher values of $S_d$.
We further observe that the increase in $\lambda$ also transforms $P(S_d)$ from
a double-peaked form to an asymmetric bell shape structure.  This stems
from the fact that the greater the $\lambda$ value, the larger the random
oscillations of $\mathcal{S}_d(\tau)$ at long times.

Further understanding of the properties of $P(S_d)$ can be gained
by noting that the values of the diagonal entropy in a certain time window
are limited to an interval of finite length and $P(S_d)$ has different shapes at
different values of $\lambda$. 
These facts, together with the results presented by one of us that concern
the modeling of the statistical distribution of Shannon entropy values \cite{WangR2020}, 
led us to fit the $P(S_d)$ values by
a beta distribution, defined as \cite{Feller1971,Johnson1995,Gupta2004}
\be \label{BtPdf}
   \varphi_B(x)=\frac{(x-S_0)^{a-1}(S_m-x)^{b-1}}{(S_m-S_0)^{a+b-1}\mathcal{B}(a,b)},
\ee
where $S_m$ denotes the maximal value of the distribution range, $a, b$ are the shape
parameters of the distribution, and $\mathcal{B}(a,b)=\int_0^1 u^{a-1}(1-u)^{b-1}du$ is the beta function. 
The cumulative distribution function of the beta distribution is given by
\be \label{BtCdf}
   \Phi_B(x)=\int_{S_0}^x \varphi_B(y)dy,
\ee
where $x$ is such that $S_0\leq x\leq S_m$.

In Fig.~\ref{Prb}, the fitted beta distribution in Eq.~(\ref{BtPdf}) and its
cumulative distribution for each case are denoted by
a red solid line in the main panel and a red dashed line in the inset.  
One can immediately identify the obvious deviation between $P(S_d)$ and the beta distribution when the value
of $\lambda$ is far away from the critical value $\lambda_c^\alpha$, in
particular for low \(\lambda\) values, as can be seen in the first and last columns of Fig.~\ref{Prb}.
However, at the critical point, with $\lambda=\lambda_c^\alpha$, the beta distribution
agrees extremely well with the numerical results, as illustrated in panels (b) and (e) of Fig.~\ref{Prb}.
To quantitatively examine the differences between $P(S_d)$ and the beta distribution, we
employ the root mean square error (RMSE), which quantifies the deviation between 
predicted and observed values \cite{Schervish2014}.
For our purpose, we consider the RMSE, denoted by $\mathcal{R}$, between the 
cumulative distribution function of the diagonal entropy and the fitted
beta distribution
\be \label{Drmse}
   \mathcal{R}=\sqrt{\left(\frac{1}{S_m-S_0}\right)\int_{S_0}^{S_m}[F(z)-\Phi_B(z)]^2dz},
\ee
where $F(z)$ and $\Phi_B(z)$ are given by Eqs.~(\ref{SdCdf}) and (\ref{BtCdf}), respectively.

In Fig.~\ref{Rmse}, we plot the dependence of $\mathcal{R}$ with $\lambda$ for
different system sizes and for  $\alpha =0.2$ and $0.4$. $\mathcal{R}$ shows an
obvious dip at the critical value $\lambda_c^\alpha$, and the minimum value
decreases for increasing system size $N$.  Therefore, the best agreement of
$P(S_d)$ with the beta distribution occurs at the critical point of the ESQPT,
as already shown in Fig.~\ref{Prb}.  Moreover, the  agreement improves
when increasing the system size, $N$.  At the critical point, we further find
that the decrease in $\mathcal{R}$ with the system size $N$ is replaced by a
tiny fluctuation around a vanishingly small value when $N>1000$,
regardless of the value of $\lambda_c^\alpha$, as shown in the insets of
Fig.~\ref{Rmse}.

The next question to address is whether the probability distribution of the
diagonal entropy, $P(S_d)$, has an universal form at the critical point of
ESQPT.  In what follows, we show that this is indeed in our case.  To this end,
we standardize the probability distribution and consider a shifted and
rescaled diagonal entropy, denoted by $\mathscr{S}_d$, defined as
\be
\mathscr{S}_d=\frac{S_d-\overline{S_d}}{\sqrt{\Sigma}},
\ee
where $\overline{S_d}=\int dS_dP(S_d)S_d$ is the averaged $S_d$ and $\Sigma=\int
dS_d P(S_d)(S_d-\overline{S_d})^2$ is the variance of $S_d$.  We now
investigate the probability distribution of $\mathscr{S}_d$, $P(\mathscr{S}_d)$,
at different values of $\lambda_c^\alpha$ and for several system sizes $N$.

Our numerical results are shown in Fig.~\ref{Updf}.  We observe that
numerical data for different $\lambda_c^\alpha$ and $N$ collapse in a single
distribution, indicating that $P(\mathscr{S}_d)$ is the universal distribution
for the ESQPT.  Moreover, the distribution $P(\mathscr{S}_d)$ can also be well
fitted by the beta distribution in Eq.~(\ref{BtPdf}), with fitting parameters
$(a,b,S_0,S_m)=(22.308,6.692,-10,3)$.  The deviations between $P(\mathscr{S}_d)$
and the fitted beta distribution are vanishingly small at different
$\lambda_c^\alpha$ and are also almost independent of the system size, $N$, as
depicted in the inset of Fig.~\ref{Updf}.  This further confirms the
universality of $P(\mathscr{S}_d)$ at the critical point of ESQPT.


\begin{figure}
    \includegraphics[width=\columnwidth]{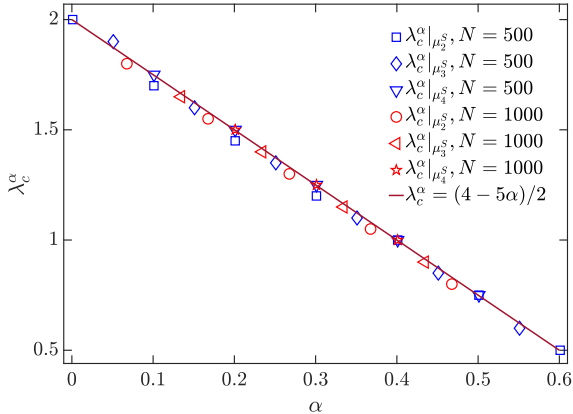}
    \caption{Critical value $\lambda_c^\alpha$, extracted from different central moments (as labelled), 
    as a function of $\alpha$ for different system sizes $N$.
    For each central moment, the location of its extreme value has been identified as the critical value $\lambda_c^\alpha$.
    The solid line denotes the analytical result, which gives by Eq.~(\ref{CriticalV}).  
    The axes in the figure are dimensionless.} 
    \label{CVm}  
  \end{figure}

\subsection*{Central moments of $P(S_d)$}

Once defined the probability distribution of the diagonal entropy, we now turn to
identify the signatures of the ESQPT in the statistical properties of $P(S_d)$, by investigating  
the central moments of $P(S_d)$.
The $n$-th central moment of $P(S_d)$ is defined as
\be
   \mu_n^S=\mathbb{E}[(S_d-\overline{S_d})^n]=
          \int_{-\infty}^{+\infty}dS_d P(S_d)(S_d-\overline{S_d})^n.
\ee
The first central moment, $\mu_1^S$, is always zero, thus we mainly focus on the moments with $n=2, 3, 4$, the variance, skewness, and kurtosis of the distribution, respectively.
These central moments provide information about the distribution shape.

In Fig.~\ref{CentralM}, we plot $\mu_2^S, \mu_3^S$ and $\mu_4^S$ as a function
of $\lambda$ for different values of the $\alpha$ control parameter   and the system size $N$.
In this figure, it is evident that the three central moments have a non-analytic behavior, with cusps 
in the neighborhood of the $\lambda_c^\alpha$ critical values.
Specifically, cusps in $\mu_2^S$ and $\mu_4^S$ display as minima that tend to zero as $N$ increases.
As the second and fourth central moments measure the fluctuations and the heaviness of the tail of a probability distribution,  
the minima values in $\mu_2^S$ and $\mu_4^S$ indicate that $P(S_d)$ has negligible fluctuations 
and becomes a light-tailed distribution in the vicinity of the ESQPT critical point, 
in accordance with the results observed in panels (b) and (e) of Fig.~\ref{Prb}.
The third central moment, $\mu_3^S$, is always less than zero, 
independently of the values of the control parameter $\alpha$ and the system size $N$.
It is known that the third central moment quantifies the distribution asymmetry.
Therefore, negative  $\mu_3^S$ values imply that the area under the left tail of $P(S_d)$ is larger than
the one under the right tail, as shown in Fig.~\ref{Prb}.
For values of the control parameter $\lambda=\lambda_c^\alpha$, the third
central moment $\mu_3^S$ shows  a cusp-like dependence toward zero, 
which is sharper for larger values of $N$.
This means that the system  $P(S_d)$ distribution has its most symmetric shape at the ESQPT critical point,
as can be seen in Figs.~\ref{Prb}(b,e) and \ref{Updf}.

The different features displayed by the central moments of $P(S_d)$ suggest that 
for a given system the critical value of $\lambda$ can be
obtained numerically from the extreme, cusp-like, values of the central moments. 
By identifying the critical point as the location of the extreme values in the central moments,
we have plotted the estimated $\lambda_c^\alpha$ as a function of $\alpha$ in Fig.~\ref{CVm}. 
We also depict in the same figure the analytical result of $\lambda_c^\alpha$
from Eq.~(\ref{CriticalV}).
As can be seen from the figure, numerical results show a good agreement with the
analytical solution,
in particular for the results from $\mu_3^S$ and $\mu_4^S$.
Moreover, the agreement can be enhanced increasing the system size.  
Therefore, we can confirm that the ESQPT has strong effects on the statistical 
properties of the probability distribution of the diagonal entropy, $P(S_d)$. 
Besides, the central moments of $P(S_d)$ can be used to 
reliably detect the critical point of the ESQPT.

\section{conclusion} \label{Sec4}

We have studied in detail the effects of the ESQPT on the dynamics and
statistics of the diagonal entropy in a quantum many-body system, the LMG model,
which undergoes an ESQPT at a certain critical energy.  We have shown that the
diagonal entropy exhibits a significant change in its time dependence as the system
goes through the critical energy of the ESQPT. Hence, the existence of an ESQPT
can be ascertained from the
calculation of the dynamics of the diagonal entropy, which also allows us to
efficiently distinguish between the different phases of the ESQPT.  To understand
the different dynamical behaviors of the diagonal entropy, we have explored the
connections between the energy dependence of $\Omega_k(E)$ [cf.~Eq.~(\ref{OmgD})] and
the dynamics of the diagonal entropy.  The results indicate the
qualitative differences in time evolution of the diagonal entropy resulting from 
 changes in $\Omega_k(E)$. In particular, at the critical energy of the
 ESQPT  the diagonal entropy follows a very particular dynamics, that can be
 traced back to the highly nontrivial cusp structures in  $\Omega_k(E)$.

The features observed in the dynamics of the diagonal entropy imply that the
ESQPT has a significant influence on  
the probability distribution of the diagonal entropy. 
We have demonstrated that the distribution of the diagonal entropy transforms
from a double peak form
to an asymmetric bell shape, once the system crosses the ESQPT. 
In particular, we have found that the distribution of the diagonal entropy can
be well described by a beta distribution at the critical point of the ESQPT.
Hence, the distribution of the diagonal entropy can be considered as a useful
tool for the ESQPT exploration.
An intriguing and remarkable result of our study is the universal behavior exhibited by the 
distribution of the diagonal entropy at the critical point of ESQPT.
We have confirmed that the distribution of the diagonal entropy values at the critical
point is independent of  both the system size and the control parameter value, and
it is in good agreement with the beta distribution.
Additionally, to examine more closely the effects of the ESQPT on the statistical properties of the diagonal entropy, we  
have analyzed the second, third, and fourth central moments of the diagonal
entropy distribution. Our results suggest that the nonanalyticities  in
the central moments make them valid probes to identify the ESQPT critical point.

The universality  of the diagonal entropy distribution at the critical point can
be traced back to the nature of the diagonal entropy time dependence in the ESQPT, which  stems from the cusps in the structure of $\Omega_k(E)$. 
We would like to emphasize that the same cusps have been found for  ESQPTs  in
various systems \cite{Perez2011,Kloc2018}, which makes us  expect that our
results are robust and hold in
other quantum many-body systems other than the LMG model, such as the Dicke model \cite{Brandes2013}, 
the kicked-top model \cite{Bastidas2014}, and the Rabi model \cite{Puebla2016}. 
A very interesting topic for future work would be a systematic study of the statistical properties of the diagonal entropy in 
different many-body systems. 
The present results pave the way to a deeper understanding of ESQPT properties and
shed light upon ESQPTs influence  on the nonequilibrium dynamics of quantum systems.
Moreover, we have also investigated the dynamical signatures of ESQPTs in
classical phase space in one of our recent work \cite{QianW2020}. Finally, the diagonal entropy
measurement  in quantum simulators  is expected to be quite efficient \cite{SunZ2020}, 
which make us believe that the obtained results could be experimentally verified in a near future.

\acknowledgments 

Q.~W. acknowledges support from the National Science Foundation of China under
Grant No.~11805165, Zhejiang Provincial Nature Science Foundation under Grant
No.~LY20A050001, and the Slovenian Research Agency (ARRS) under the Grants
No.~J1-9112 and No.~P1-0306. FPB thanks the support of the Consejer\'{\i}a de Conocimiento, Investigaci\'on y Universidades, Junta de
Andaluc\'{\i}a and European Regional Development Fund (ERDF) through projects 
SOMM17/6105/UGR and UHU-1262561 and the support of the Ministerio de Ciencia,
Innovaci\'on y Universidades through project PID2019-104002GB-C21.  Computing
resources supporting this work were partly provided by the CEAFMC and
Universidad de Huelva High Performance Computer (HPC@UHU) located in the Campus
Universitario el Carmen and funded by FEDER/MINECO project UNHU-15CE-2848.

\bibliographystyle{apsrev4-1}
\bibliography{DEesqpt}

\end{document}